\documentclass{article}
\usepackage{cite}

\def\fracd#1#2{{\displaystyle\frac{#1}{#2}}}
\def\sumd{\displaystyle\sum}

\begin{document}

\title{On Computation of Combined IVS EOP Series}
\author{Zinovy Malkin \\ Central (Pulkovo) Astronomical Observatory RAS, St.~Petersburg, Russia}
\date{~}
\maketitle

\footnotetext{Presented at 15th Working Meeting on European VLBI for Geodesy
and Astrometry, Barcelona, Spain, Sep 07-08, 2001}

\begin{abstract}
Three topics related to computation of the combined IVS EOP series are
discussed.  The first one is the consistency of the VLBI
EOP series with the IERS reference frames ITRF and ICRF.
Not all IVS analysis centers use ITRF/ICRF as reference system for
EOP solutions.  Some of them realize global solution for simultaneous
determination of CRF, TRF and EOP with no-net-rotation constrains
w.r.t. ITRF/ICRF.  Analysis shows
that such a method can hardly provide consistency of computed EOP series
with the IERS reference frames with sufficient accuracy.
Direct use of ITRF and ICRF
for computation of EOP series submitted to the IERS seems preferable.
Second problem is the long-time stability of the IVS EOP series.
Analysis of yearly biases w.r.t. C04 and NEOS is presented.
A possible ways are proposed
to save long time stability of the combined IVS EOP series.
At last, various strategies of computation of weighted mean value are
considered.
It's shown that usual methods used for this purpose do not provide
satisfactory result for the error of the mean.
A new method is proposed.
\end{abstract}

\section{Introduction}

The IVS combined EOP series computed at the IVS Analysis Coordinator
Office located at the Geodetic Institute of the University of Bonn
is available beginning from the end of 2000.  Analysis of this series
routinely provided by the IERS EOP Product Center at the Paris Observatory
shows that its accuracy is better than accuracy of individual solutions
provided by the IVS Analysis Centers.  However, some topics related
to the quality of the IVS combined EOP series seems to be investigated
more carefully.  This paper is intended to consider the following points:
\begin{itemize}
\item Consistency of the VLBI EOP series with the IERS reference frames
ITRF and ICRF.
\item Systematic stability of the VLBI EOP series.
\item Computation of weighted mean values.
\end{itemize}

\section{Consistency of IVS EOP series with IERS reference systems}

According to the IVS Terms of Reference, IVS serves as the VLBI Technique
Center for IERS.  In turn, the IERS Terms of Reference said that one
of the IERS primary objectives is providing Earth orientation parameters (EOP)
required to transform between ICRF and ITRF.  It is supposed that after
completion of new IERS structure the IERS EOP product will be computed
combining several EOP series delivered by the IERS Technique Centers one
of which is the IVS.  So, the evident goal of the IVS is computation of
the combined IVS EOP series providing the transformation parameters
between ITRS and ICRS.

However, not all IVS analysis centers use ITRF/ICRF as reference system for
EOP solutions.
Some of them realize global solution for simultaneous
determination of CRF, TRF and EOP.
To tie a global solution to IERS reference
frames no-net-rotation constrains w.r.t. ITRF and ICRF are usually applied.
The question is can such a method provide the consistency of VLBI
EOP series with ITRF/ICRF with required accuracy?

Usually global VLBI solution is made using all available sessions
and application of no-net-rotation provides zero translation and
rotation of full set of stations and radio sources w.r.t. ITRF and ICRF.
However, commonly speaking, it is not the case for the subset of stations
participating in a particular session.
Therefore EOP estimate obtained from processing of a session observations
may systematically differ from ITRF/ICRF.

Besides, number of observations for stations and sources differ very much.
Table~\ref{tab:obs_stat} shows statistics of observations for
stations and sources for all sessions and for NEOS-A ones.
One can see that in fact the NEOS-A EOP series used in the IVS combined
solution is practically defined by subset of 8 stations and 66 radio sources.

\begin{table}[ht]
\centering
\caption{Statistics of observations (Nobs in thousands).}
\label{tab:obs_stat}
\medskip
\begin{tabular}{|cc|cc|cc|cc|}
\hline
\multicolumn{4}{|c|}{NEOS-A} & \multicolumn{4}{|c|}{All sessions} \\
\hline
\multicolumn{2}{|c|}{Stations} & \multicolumn{2}{|c|}{Sources} &
\multicolumn{2}{|c|}{Stations} & \multicolumn{2}{|c|}{Sources} \\
\hline
Nsta & Nobs & Nsou & Nobs  & Nsta & Nobs & Nsou & Nobs \\
\hline
5 & 100--200 & ~11 & 10--30 & 15 & 100-700 & ~14 & 50--200 \\
3 & 50--100  & ~25 & 5--10  & 32 & 10--100 & ~49 & 10--50  \\
7 & 1--6     & ~39 & 1--5   & 52 & 1--10   & ~76 & 1--10   \\
6 & $<1$     & ~95 & 0.1--1 & 50 & $<1$    & 299 & 0.1--1  \\
  &          & 177 & $<0.1$ &   &          & 341 & $<0.1$  \\
\hline
\end{tabular}
\end{table}

Let us see how close is the tie between different subsets of station and
source catalogs with ITRF/ICRF.
We use results of the USNO9903 global solution as most fresh publicly
available one.
Tables~\ref{tab:usno_icrf} and \ref{tab:usno_itrf} present transformation
parameters between USNO solution and ITRF(ICRF) for all common stations
(sources), for all stations (sources) participating in the NEOS-A program,
and for most frequently observing stations (observed sources).
These data show that the transformation parameters including ones
defining EOP system are not equal to zero and differ for various subsets
of stations and sources.  This mean that EOP system is not correspond
to ITRF/ICRF and differ for various observational programs.

Therefore, CRF and TRF realization obtained from a global VLBI solution
can hardly provide consistency of computed EOP series
with ITRF/ICRF with sufficient accuracy.
Direct use of the ITRF and ICRF
for computation of EOP series submitted to the IERS seems preferable.

This does not mean indeed that the IVS Analysis centers should not compute
global solutions.
The reasonable strategy may be using individual CRF and TRF realizations
for improving the IERS reference frames,
and further using ICRF and ITRF for regular EOP computation.
This strategy provides consistency
between VLBI EOP series and the IERS reference frames and
makes individual VLBI EOP series more homogeneous that
allows to simplify combination procedure and improve quality of
the IVS combined product.

\begin{table}
\parbox[t]{79mm}{
\centering
\caption{Transformation parameters between the TRF realizations
USNO9903 and ITRF97 at epoch 1997.0 for different number of stations.}
\label{tab:usno_itrf}
\medskip
\begin{tabular}{|l|r|r|r|}
\hline
N sta     &     85~   &    20~   &     8~~  \\
\hline
 T1, mm   &   --0.1   &  --0.9   &  --2.2   \\
$\sigma$  &     0.8   &    1.6   &    1.9   \\
\hline
 T2, mm   &   --1.6   &  --1.4   &  --1.3   \\
$\sigma$  &     0.8   &    1.7   &    2.0   \\
\hline
 T3, mm   &   --0.6   &  --0.4   &  --0.3   \\
$\sigma$  &     0.8   &    1.6   &    1.8   \\
\hline
 D, ppb   &     1.7   &    1.5   &    0.8   \\
$\sigma$  &     0.1   &    0.2   &    0.2   \\
\hline
\hline
 R1, mas  &  --0.06   & --0.05   & --0.03   \\
$\sigma$  &    0.03   &   0.07   &   0.08   \\
\hline
 R2, mas  &    0.01   &   0.02   &   0.08   \\
$\sigma$  &    0.03   &   0.05   &   0.06   \\
\hline
 R3, mas  &    0.00   &   0.03   &   0.00   \\
$\sigma$  &    0.03   &   0.05   &   0.06   \\
\hline
\end{tabular}
}
\hspace{4mm}
\parbox[t]{79mm}{
\centering
\caption{Transformation parameters between the CRF realizations
USNO9903 and ICRF-Ext.1 for different number of sources.}
\label{tab:usno_icrf}
\medskip
\begin{tabular}{|l|r|r|r|}
\hline
N sou            &    626~~  &   303~~  &   66~~~  \\
\hline
A$_1$, mas       &   0.029   &   0.022  &   0.013  \\
$\sigma$         &       9   &       9  &       7  \\
\hline
A$_2$, mas       &   0.027   &   0.018  &   0.013  \\
$\sigma$         &       9   &       8  &       7  \\
\hline
A$_3$, mas       & --0.018   & --0.016  & --0.013  \\
$\sigma$         &       9   &       9  &      12  \\
\hline
D$_\alpha$, mas  & --0.001   & --0.001  &       0  \\
$\sigma$         &       0   &       0  &       0  \\
\hline
D$_\delta$, mas  &       0   &       0  & --0.001  \\
$\sigma$         &       0   &       0  &       0  \\
\hline
B$_\delta$, mas  &   0.054   &   0.042  &   0.086  \\
$\sigma$         &       9   &      10  &      12  \\
\hline
\end{tabular}
}
\end{table}

Now the IERS and main space geodesy services are in the process of
moving from ITRF97 to ITRF2000.
What systematic changes in the VLBI EOP series can we expect?
Table~\ref{tab:itrf} shows the results of comparison
between ITRF97 and ITRF2000 for different subsets
of VLBI stations.

\begin{table}
\centering
\caption{Transformation parameters between ITRF97 and ITRF2000
at epoch 1997.0
for different number of VLBI stations (98, 20, 8), nominal transformation
parameters defined by the IERS ($P_0$) and systematic differences between
EOP series computed with ITRF97 and ITRF2000 with the OCCAM package.}
\label{tab:itrf}
\medskip
\begin{tabular}{|l|r|r|r|r|r|}
\hline
N sta      &    98~ &    20~ &    8~~ &  $P_0$~ & OCCAM \\
\hline
 T1, mm    &    6.9 &    7.8 &    7.8 &    6.7  &   \\
$\sigma$   &    0.5 &    0.9 &    1.4 &         &   \\
\hline
 T2, mm    &    3.9 &    3.9 &    3.6 &    6.1  &   \\
$\sigma$   &    0.5 &    1.0 &    1.6 &         &   \\
\hline
 T3, mm    & --20.2 & --20.1 & --21.3 & --18.5  &   \\
$\sigma$   &    0.5 &    0.9 &    1.4 &         &   \\
\hline
 D, ppb    &    1.5 &    1.3 &    0.9 &   1.55  &   \\
$\sigma$   &    0.1 &    0.1 &    0.2 &         &   \\
\hline
 R1, mas   & --0.17 & --0.19 & --0.20 &   0.0   & --0.15~~~ \\
$\sigma$   &   0.02 &   0.04 &   0.06 &         &   0.02~~~ \\
\hline
 R2, mas   &   0.01 & --0.01 & --0.01 &   0.0   &   0.09~~~ \\
$\sigma$   &   0.02 &   0.03 &   0.05 &         &   0.02~~~ \\
\hline
 R3, mas   & --0.01 & --0.03 & --0.03 &   0.0   & --0.07~~~ \\
$\sigma$   &   0.02 &   0.02 &   0.04 &         &   0.01~~~ \\
\hline
&&&&&\\[-1em]
\.R3, mas/y&        &        &        & --0.02  & --0.03~~~ \\
$\sigma$   &        &        &        &         &   0.02~~~ \\
\hline
\end{tabular}
\end{table}

We also compared the EOP series computed at the IAA with ITRF97 and
ITRF2000.  The result is shown in Table~\ref{tab:itrf}.
The computation was made with three radio source catalogues ICRF-Ext.1,
RSC(IAA)99R02 and RSC(IAA)01R02, and no meaningful systematic differences
between EOP series computed with ITRF97 and ITRF2000 was found.

Although ITRF2000 was constructed in such a way that no rotation
w.r.t. ITRF97 was introduced, substantial value of rotation angle R2
(which corresponds to Y pole coordinate) is found both in direct
comparison of two coordinate systems and in the result of EOP computation.

In conclusion, it's important to mention that errors of inconsistency
between EOP series and terrestrial and celestial reference frames are
systematic ones and even their relatively small values can be
substantial.

\section{Long-time stability of IVS EOP series}

Obviously, one of the main goal of maintenance of the IVS combined
products is to provide systematically stable IVS EOP series.
It is especially important now because the new IERS organization
envisages computation of the IERS combined EOP series practically
of the three series VLBI, GPS and SLR, and importance of each of
them is very large.
Several factors make this task difficult and in the first place it is
instability of individual series.  The main reason for that are:
\begin{itemize}
\item Using individual periodically updated TRF and CRF realizations.
As shown in the previous section these realization are not tied to
the unique (IERS) reference frames with sufficient accuracy and, in fact,
every VLBI session yields EOP estimates in its own system.
\item Change in systematic errors of EOP series after modification
of models, algorithms and software.
\item Change of set of contributed VLBI Analysis Centers.  Besides
a difference in used reference systems, each EOP series has its own
systematic peculiarities.
\item Change of network configuration.  This is well established fact,
and it is not quite clear how to handle it properly.  For instance, we
can mention the problem of joining 9-year IRIS-A and 8-year NEOS-A
programs to avoid EOP jump directly affected results of determination
of 18.6-year nutation term.
\end{itemize}

For listed above and other reasons the VLBI EOP series show long-time
instability.  To investigated this effect we use five VLBI EOP series
BKG00001, GSF2000A, IAAO9907, SPU00001, USN99003 over a 7-year interval
from May 1993 till April 2000 (NEOS-A data only).
The whole 7-year interval was split in 7 one-year ones and each series
was compared with combined C04 and NEOS series at these one-year intervals.
During computation six parameters of systematic differences between VLBI
and combined series were estimated for every year.
These are: bias, rate, amplitude
of sine and cosine of annual term and semiannual terms.
In such a way
we obtained seven values for each of six parameters of model of systematic
errors for each VLBI series.
The final step of this analysis was the computation of RMS values from
seven epochs.
Such a approach to investigation of long-time stability is
analogous to a method used at the Paris observatory during computation
of the IERS combined products.
Result of analysis of yearly biases is presented in
Table~\ref{tab:long_time_stab}.

\begin{table}[ht]
\centering
\caption{Long-time stability of IVS EOP series (NEOS-A):
statistics of yearly bias relative to the IERS C04 and NEOS
combined series (7 years 1993.3--2000.3):
bias, rate - result of approximation of yearly bias series
by linear trend, rms - rms of residuals after removing trend.}
\label{tab:long_time_stab}
\medskip
\tabcolsep=4.5pt
\begin{tabular}{|l|l||rrrrr||rrrrr|}
\hline
&& \multicolumn{5}{|c||}{C04} & \multicolumn{5}{c|}{NEOS} \\
\cline{3-12}
EOP && BKG & GSF & IAA & SPU & USN & BKG & GSF & IAA & SPU & USN \\
\hline
&&&&&&&&&&& \\
X     & bias & $ 0.064$ & $-0.088$ & $-0.126$ & $-0.074$ & $-0.096$ & $ 0.080$ & $-0.071$ & $-0.111$ & $-0.058$ & $-0.081$ \\
mas   & rate & $ 0.011$ & $ 0.015$ & $-0.002$ & $ 0.002$ & $ 0.009$ & $ 0.034$ & $ 0.037$ & $ 0.021$ & $ 0.024$ & $ 0.032$ \\
      & rms  & $ 0.025$ & $ 0.026$ & $ 0.025$ & $ 0.027$ & $ 0.023$ & $ 0.034$ & $ 0.028$ & $ 0.041$ & $ 0.034$ & $ 0.036$ \\
&&&&&&&&&&& \\
Y     & bias & $-0.249$ & $ 0.015$ & $-0.065$ & $-0.030$ & $-0.034$ & $-0.269$ & $-0.004$ & $-0.078$ & $-0.043$ & $-0.048$ \\
mas   & rate & $ 0.064$ & $-0.003$ & $ 0.052$ & $ 0.049$ & $ 0.043$ & $ 0.064$ & $-0.002$ & $ 0.056$ & $ 0.052$ & $ 0.046$ \\
      & rms  & $ 0.040$ & $ 0.031$ & $ 0.049$ & $ 0.043$ & $ 0.044$ & $ 0.052$ & $ 0.041$ & $ 0.054$ & $ 0.050$ & $ 0.051$ \\
&&&&&&&&&&& \\
UT1   & bias & $ 0.101$ & $-0.020$ & $-0.037$ & $-0.232$ & $-0.041$ & $ 0.163$ & $ 0.038$ & $ 0.025$ & $-0.170$ & $ 0.024$ \\
0.1 ms& rate & $ 0.018$ & $-0.056$ & $-0.014$ & $ 0.018$ & $-0.003$ & $ 0.010$ & $-0.062$ & $-0.016$ & $ 0.014$ & $-0.008$ \\
      & rms  & $ 0.026$ & $ 0.020$ & $ 0.022$ & $ 0.018$ & $ 0.019$ & $ 0.044$ & $ 0.048$ & $ 0.055$ & $ 0.062$ & $ 0.050$ \\
&&&&&&&&&&& \\
dPsi  & bias & $-0.066$ & $-0.049$ & $-0.061$ & $ 0.030$ & $ 0.080$ & $ 0.066$ & $ 0.087$ & $ 0.072$ & $ 0.165$ & $ 0.212$ \\
mas   & rate & $-0.028$ & $-0.014$ & $ 0.013$ & $ 0.002$ & $-0.025$ & $-0.009$ & $ 0.006$ & $ 0.027$ & $ 0.016$ & $-0.009$ \\
      & rms  & $ 0.087$ & $ 0.083$ & $ 0.031$ & $ 0.054$ & $ 0.060$ & $ 0.056$ & $ 0.040$ & $ 0.049$ & $ 0.055$ & $ 0.031$ \\
&&&&&&&&&&& \\
dEps  & bias & $ 0.001$ & $-0.008$ & $ 0.047$ & $-0.021$ & $-0.043$ & $ 0.037$ & $ 0.027$ & $ 0.085$ & $ 0.021$ & $-0.002$ \\
mas   & rate & $ 0.006$ & $-0.005$ & $ 0.007$ & $-0.012$ & $-0.009$ & $ 0.013$ & $ 0.002$ & $ 0.015$ & $-0.003$ & $-0.001$ \\
      & rms  & $ 0.020$ & $ 0.009$ & $ 0.031$ & $ 0.031$ & $ 0.016$ & $ 0.027$ & $ 0.016$ & $ 0.039$ & $ 0.042$ & $ 0.022$ \\
\hline
\end{tabular}
\end{table}

Obviously, this analysis cannot be fully objective because it depends
on details of combination procedure (systematic corrections, weights, etc.)
used during computation of C04 and NEOS series.
One can see that differences between the left and the right
parts of Table~\ref{tab:long_time_stab} is sometimes quite large, especially
for UT1-UTC.
Maybe using IVS combined EOP series for such a analysis would be
preferable when it will have sufficient time span.

The results of analysis presented here and in the previous section
confirm well known fact that each EOP series has own systematic
errors and these errors are not stable at the required level of accuracy.
Therefore it seems very important to develop appropriate strategy
for computation of the IVS combined product to provide make its
systematic stability.
We would like to propose for discussion a possible strategy to keep
long-time systematic stability of the IVS EOP combined series.
This strategy includes the following steps.

\begin{enumerate}
\item Computation of the ``reference'' EOP series $EOP_0$ as the mean
of existing long-time NEOS-A series
{\it fixed at epoch of computation}
with weights depending on long-time stability.  Input series should be
transformed to uniform TRF/CRF (preferably the IERS ones)
as accurate as possible.
\item Using systematic corrections to individual series
$$dEOP_i=EOP_0-EOP_i$$
derived from comparison with the reference series in further computations.
\item When an AC$_i$ updates EOP series new systematic correction
can be computed as
$$(EOP_{i,old}-EOP_{i,new})+dEOP_i \ .$$
\item When a new EOP series of a new AC is to be included in the
IVS combination systematic correction to that series will be
$$dEOP_j=EOP_0-EOP_j \ .$$
\item Periodical update of the reference series,
{\it e.g.} when new ITRF or ICRF realization is accepted.
Evidently, in such a case, appropriate care of careful tie
between the new and the old reference series must be taken.
\end{enumerate}

A separate problem is the transformation of EOP obtained on different networks
to the reference series.  However, hopefully improvement of ITRF and
models of VLBI observations will eliminate this problem in the future.

\section{Computation of weighted mean}

Computation of the weighted mean of several estimates is usually the final
step in each EOP (and all others) combining procedure.
Let we have $n$ values $x_i$ with associated errors $s_i$, $i=1 \ldots n$.
Then we have a well known statistics \cite{Brandt75,Bevington69}
$$
p_i = \fracd{1}{s_i^2} \ , \qquad
p = \sumd_{i=1}^n {p_i} \ , \qquad
x = \fracd{\sumd_{i=1}^n  p_i x_i}{p} \ ,
$$
$$
H = \sumd_{i=1}^n {p_i (x_i-x)^2}
\ = \ \sumd_{i=1}^n {\left[\fracd{(x_i-x)}{s_i}\right]^2} \ , \quad
\chi^2/dof = \fracd{H}{n-1} \ ,
$$
\noindent where $x$ is a estimate of the mean value.
The question is how to estimate error $\sigma$ of the mean?
Two classical approaches are:

Maximum likelihood approach if $\sigma_i$ are considered as
absolute magnitudes of errors in $x_i$:
$$
\sigma_1 = \fracd{1}{\sqrt{p}} \ .
$$

Least squares approach if $\sigma_i$ are considered
as relative values of errors in $x_i$ and error of unit weight must
be estimated from data itself:
$$
\sigma_2
= \sqrt{\fracd{\sumd_{i=1}^n {p_i (x_i-x)^2}}{p\,(n-1)}}
\ = \ \sqrt{\fracd{H}{p\,(n-1)}}
\ = \  \sigma_1\,\sqrt{\fracd{H}{n-1}} \ .
$$

It is easy to see that $\sigma_1$ depends only on a priori errors in
averaged values $x_i$ and $\sigma_1$ depends only on the scatter of $x_i$.
Theoretically, solution of problem of choice between $\sigma_1$ and $\sigma_2$
depend on whether the scatter of $x_i$ is a result of random errors or
there are systematic differences between estimates $x_i$.
Obviously, both effect are present in most of practical applications.

That is a known problem in data processing and no rigorous solution
is proposed.  However some practical ways to handle it were considered
in literature.  Evidently, the most statistically substantial approach
was made in \cite{Brandt75,Rosenfeld67}.  According to this approach
chi-square criteria is used to decide if the scatter of $x_i$ is result of
random errors, and error of the mean $x$ is computed as
$$
\sigma_3 = \left\{
\begin{array}{ll}
\sigma_1, & \ \mbox{if $H \le \chi^2(Q,n-1)$} \ , \\[1ex]
\sigma_2, & \ \mbox{if $H > \chi^2(Q,n-1)$} \ ,
\end{array}
\right.
$$
where $Q$ is a fiducial probability.
Some other practical algorithms of choice between $\sigma_1$ and $\sigma_2$
were proposed too.

However, in practice, values of $\sigma_1$ and $\sigma_2$ may differ by
several times.  It leads to instability of $\sigma$ estimate.
Table~\ref{tab:stat_examples} shows some numerical
examples of computation of weighted mean of two data points and its error
(to compute $\sigma_3$ we use $Q$=99\% which corresponds $\chi^2$(0.99,1)=6.63).
One can see that no one value of $\sigma_1$, $\sigma_2$, $\sigma_3$
provides a satisfactory estimate of $\sigma$.
Moreover, value of $\sigma_3$ depends not only on data sample
$\{x_i,\, s_i\}$ but also on subjective choice of $Q$.

\begin{table}[ht]
\centering
\caption{Numerical examples of computation of weighted mean (see explanation
in text).}
\label{tab:stat_examples}
\medskip
\begin{tabular}{|r|rr|r|r|r|rrrr|}
\hline
No & $x_1$ & $x_2$ & $s_{1,2}$ & \multicolumn{1}{c|}{$x$} & \multicolumn{1}{c|}{$H$} &
\multicolumn{1}{c}{$\sigma_1$} & \multicolumn{1}{c}{$\sigma_2$} &
\multicolumn{1}{c}{$\sigma_3$} & \multicolumn{1}{c|}{$\sigma_4$} \\
\hline
&&&&&&&&& \\
 1 &  1.0 & 1.0  &  0.5 &  1.0 &    0.00 &     0.354 & 0.000 & 0.354 & 0.354 \\
&&&&&&&&& \\
 2 &  1.0 &  2.0 &  0.1 &  1.5 &   50.00 &     0.071 & 0.500 & 0.500 & 0.505 \\
 3 &      &      &  0.2 &      &   12.50 &     0.141 & 0.500 & 0.500 & 0.520 \\
 4 &      &      &  0.3 &      &    5.56 &     0.212 & 0.500 & 0.212 & 0.543 \\
 5 &      &      &  0.5 &      &    2.00 &     0.354 & 0.500 & 0.354 & 0.612 \\
 6 &      &      &  1.0 &      &    0.50 &     0.707 & 0.500 & 0.707 & 0.866 \\
 7 &      &      &  2.0 &      &    0.12 &     1.414 & 0.500 & 1.414 & 1.500 \\
&&&&&&&&& \\
 8 & 10.0 & 20.0 &  0.1 & 15.0 & 5000.00 &     0.071 & 5.000 & 5.000 & 5.000 \\
 9 &      &      &  0.5 &      &  200.00 &     0.354 & 5.000 & 5.000 & 5.012 \\
10 &      &      &  1.0 &      &   50.00 &     0.707 & 5.000 & 5.000 & 5.050 \\
11 &      &      &  2.0 &      &   12.50 &     1.414 & 5.000 & 5.000 & 5.196 \\
12 &      &      &  3.0 &      &    5.56 &     2.121 & 5.000 & 2.121 & 5.431 \\
13 &      &      &  5.0 &      &    2.00 &     3.536 & 5.000 & 3.536 & 6.124 \\
14 &      &      & 10.0 &      &    0.50 &     7.071 & 5.000 & 7.071 & 8.660 \\
15 &      &      & 20.0 &      &    0.12 &    14.142 & 5.000 &14.142 &15.000 \\
&&&&&&&&& \\
16 & 10.0 & 10.0 & 1.0  & 10.0 &   0.00  &     0.707 & 0.000 & 0.707 & 0.707 \\
17 & 10.0 & 11.0 &      & 10.5 &   0.50  &     0.707 & 0.500 & 0.707 & 0.866 \\
18 & 10.0 & 12.0 &      & 11.0 &   2.00  &     0.707 & 1.000 & 0.707 & 1.225 \\
19 & 10.0 & 13.0 &      & 11.5 &   4.50  &     0.707 & 1.500 & 0.707 & 1.658 \\
20 & 10.0 & 14.0 &      & 12.0 &   8.00  &     0.707 & 2.000 & 2.000 & 2.121 \\
21 & 10.0 & 15.0 &      & 12.5 &  12.50  &     0.707 & 2.500 & 2.500 & 2.598 \\
22 & 10.0 & 16.0 &      & 13.0 &  18.00  &     0.707 & 3.000 & 3.000 & 3.082 \\
23 & 10.0 & 17.0 &      & 13.5 &  24.50  &     0.707 & 3.500 & 3.500 & 3.571 \\
&&&&&&&&& \\
\hline
\end{tabular}
\end{table}

After many experiments with test data we decided in favor of simple formula
$$
\sigma_4=\sqrt{\sigma_1^2+\sigma_2^2} =
\sqrt{\fracd{1}{p}\left(1+\fracd{H}{n-1}\right)} \ ,
$$
which can be called ``combined'' approach.
The last column of Table~\ref{tab:stat_examples} shows that such a approach
can provide stable and realistic estimate of error of the mean.

More detailed consideration of this topic is given in \cite{Malkin01f}.

\section{Conclusions}

Results of this study allow to make the following conclusions:
\begin{itemize}
\item Procedure of computation of the IVS combined EOP series
must be ``absolute'', i.e. independent on any reference, e.g. IERS, series.
Otherwise details of combination procedure used during computation of
``external'' reference series (systematic corrections, weights) will
affect the results of analysis.
\item It seems preferable to use ITRF and ICRF by the all IVS Analysis Centers
for computation of VLBI EOP series submitted to the IVS and IERS.
Using individual TRF/CRF lead to difficulties in interpretation of results.
Usual procedure of determination of systematic differences between EOP series
provides correction only for ``global'' orientation between TRF/CRF.
But, as it was shown above transformation parameters between individual
TRF (CRF) realizations depend on sub-set of stations (sources) used for
comparison.  This means that, commonly speaking, every session produce
EOP in its own system, which makes it difficult to transform an individual
EOP series to ITRF/ICRF with sufficient accuracy.
\item A reference EOP series based on IVS combined solution for fixed set
of individual solutions can be used to save the long-time stability.
Also, it is important to develop appropriate strategy to include new
or updated solutions in the IVS combination, e.g. using strategy
proposed in this paper.
\item Weighting of individual series depending on their long-time
stability seems useful for improvement of long-time stability of
the IVS combined EOP series.
\item Proposed method of computation of weighted mean EOP can
be used to account for both formal error and scatter.
\end{itemize}


\begin{thebibliography}{99}

\bibitem{Brandt75}
Brandt S. Statistical and Computational Methods in Data Analysis. 1970.

\bibitem{Bevington69}
Bevington P. R.
Data reduction and error analysis for the physical sciences.
McGraw--Hill Book Company, USA, 1969.

\bibitem{Rosenfeld67}
Rosenfeld A. H., Barbero-Galtieri A., Podolski~W.~J., Price~L.~R.,
Soding~P., Wohl~C.~G.
Data on Particles and Resonant States.
Rev. Mod. Phys., 1967, {\bf 39}, No 1, 1--51.

\bibitem{Malkin01f}
Malkin Z. M. On Computation of Weighted Mean. Comm. IAA, 2001, No 137.
(in Russian)

\end{thebibliography}
\end{document}